\RequirePackage{fixltx2e}
\documentclass[10pt,a4paper,aip,jcp,reprint,floatfix]{revtex4-1}

\usepackage[utf8]{inputenc}
\usepackage{xcolor}
\usepackage{hyperref}
\hypersetup{bookmarks, bookmarksnumbered}
\hypersetup{colorlinks, citecolor=blue}
\usepackage{amsmath}
\usepackage{amsfonts}
\usepackage{xspace}
\usepackage{graphicx}
\DeclareGraphicsExtensions{.pdf,.jpg}
\newlength{\singlefigure}
\newlength{\doublefigure}
\newcommand{\unit}[1]{\ensuremath{ \mathbf{\hat{#1}} }} %unit vectors

\newcommand{\kb}{\ensuremath{k_{\mathrm{B}}}\xspace} %Boltzmanns constant
\newcommand{\Tstar}{\ensuremath{T^{*}}\xspace} %reduced temperature
\newcommand\acie{Angew. Chem. Int. Ed.}
\newcommand\bp{Biopolymers}
\newcommand\bpj{Biophys. J.}
\newcommand\cpc{Comp. Phys. Comm.}
\newcommand\ieeetr{IEEE Trans. Robotics}
\newcommand\jcp{J. Chem. Phys.}
\newcommand\jpcb{J. Phys. Chem. B}
\newcommand\jpcm{J. Phys.:Cond. Matt.}
\newcommand\mmrc{Macromol. Rapid Comm.}
\newcommand\nat{Nature}
\newcommand\nmat{Nature Materials}
\newcommand\pccp{Phys. Chem. Chem. Phys.}
\newcommand\pb{Phys. Biol.}
\newcommand\pnasu{Proc. Nat. Acad. Sci. USA}
\newcommand\pre{Phys. Rev. E}
\newcommand\prl{Phys. Rev. Lett.}
\newcommand\sci{Science}
\newcommand\sm{Soft Matter}
\newcommand\tbs{Trends Biochem. Sci.}
\newcommand\vir{Virology}
\newcommand\green[1]{#1}

\begin{document}

\title{Design Strategies for Self-Assembly of Discrete Targets}
\author{Jim Madge}
\author{Mark A. Miller}
\email{m.a.miller@durham.ac.uk}
\affiliation{
  Department of Chemistry,\
  Durham University,\
  South Road,\
  Durham,\
  DH1 3LE,\
  United Kingdom
}
\date{\today}

\begin{abstract}

Both biological and artificial self-assembly processes can take place by a range
of different schemes, from the successive addition of identical building blocks,
to hierarchical sequences of intermediates, all the way to the fully addressable
limit in which each component is unique.  In this paper we introduce an
idealized model of cubic particles with patterned faces that allows
self-assembly strategies to be compared and tested.  We consider a simple
octameric target, starting with the minimal requirements for successful
self-assembly and comparing the benefits and limitations of more sophisticated
hierarchical and addressable schemes.  Simulations are performed using a hybrid
dynamical Monte Carlo protocol that allows self-assembling clusters to rearrange
internally while still providing Stokes--Einstein-like diffusion of aggregates
of different sizes.  Our simulations explicitly capture the thermodynamic,
dynamic and steric challenges typically faced by self-assembly processes,
including competition between multiple partially-completed structures.
Self-assembly pathways are extracted from the simulation trajectories by a fully
extendable scheme for identifying structural fragments, which are then assembled
into history diagrams for successfully completed target structures.  For the
simple target, a one-component assembly scheme is most efficient and robust
overall, but hierarchical and addressable strategies can have an advantage under
some conditions if high yield is a priority.

\end{abstract}

\maketitle

\section{Introduction}
\label{sec:Introduction}

A vast range of physical phenomena have been legitimately described as a form of
`self-assembly'.  The uniting features of these processes provide a
minimal definition of self-assembly with just two
criteria: that an ordered structure emerges from a state where the components
were either highly disorganized or widely separated, and that no detailed external
influence is applied to make the process of organization take place.  The
latter requirement implies that self-assembly is a spontaneous process, driven
by the energetic interactions between the particles\cite{Bishop09b} and by
the entropy of the system as a
whole.\cite{Frenkel15a,vanAnders14a}  Information about a target structure is therefore
implicitly encoded in its constituent building blocks and in the medium in
which the building blocks exist.
\par
In soft matter, there is great scope for synthesizing macromolecular and colloidal
building blocks with bespoke shapes and interactions.\cite{Sacanna13a}
The continually advancing experimental possibilities open up the attractive
prospect of approaching nanoscale self-assembly from the bottom
up\cite{Cademartiri15a}---in other words, of exerting
detailed control over the final structure, and even
over the pathway by which it is reached.\cite{Rogers15a}
Sets of principles are beginning to be established to provide guidance on the
design of building blocks and the background medium
for targeted self-assembly.\cite{Escobedo14a,Jacobs15b}
\par
When considering how to self-assemble a particular target, a range of strategies
is available.  Some of the earliest studies of self-assembly were inspired by
the remarkable ability of certain icosahedral virus capsids to form from a
precise number of copies---an integer multiple $T$ of 60---of the same
protein.\cite{Bancroft67a}  Nature's `strategy' here is one of economy: a
one-component construction makes minimal demands on the limited resources of a
system as small as a virus.  To assemble multiple copies of a monodisperse,
discrete capsid from a suspension of identical protein building blocks, viruses
face a number of generic obstacles encountered elsewhere in self-assembly: the
system must approach a thermodynamically stable state while avoiding amorphous
aggregation, allowing imperfections in assembly to be corrected, and preventing
partially completed structures from starving each other of building
blocks.\cite{Hagan06a,Zandi06a,Sweeney08a,Fejer09a,Rapaport10a} The high
symmetry of the icosahedron clearly plays an important role in making
one-component assembly possible.  However, capsids with $T>1$ go even further
than exploiting the equivalence of sites in high-symmetry structures, since the
local environments of individual proteins in such capsids are not identical but
merely similar.  Some $T=3$ capsids, for example, contain identical proteins in
three distinguishable local environments.  This phenomenon of
quasi-equivalence\cite{Caspar62a} presses the efficiency of the one-component
strategy to the limits.
\par
Much more recently, it has been shown that self-assembly can be achieved by a
strategy that is quite the opposite of the minimal case of viruses.  In fully
addressable self-assembly, each building block is
programmed\cite{Macfarlane13a,Jones15a} to occupy a specific site in the target
structure.  To achieve this level of precision, each building block must be
unique and be encoded with enough information to guide it to the desired
location without interference from building blocks that are not near it in the
target structure.  Such selective interactions can be realized by exploiting the
specificity of nucleotide base pairs.\cite{Mirkin96a,Alivisatos96a}  Structural
elements can be created by folding short sequences of single-stranded DNA into
tiles\cite{Wei12a} or bricks\cite{Ke12a}, which then self-assemble into precise
structures that may have a thousand different components.  Other DNA-based
schemes for self-assembly, notably DNA origami,\cite{Rothemund06a} also rely on
the addressability of DNA by using sections of a longer scaffold strand to
specify the locations for short staple strands that hold the structure together.
\par
Lying between the minimal and fully-addressable limits there is the possibility
of a hierarchical strategy to self-assembly.  Such a multiple-step approach is
intuitively appealing if the target itself has a modular structure that can be
decomposed into subunits.  Hierarchical assembly has been observed and exploited
in a wide range of systems, including DNA polyhedra constructed in two stages
from tiles of multiple single strands, \cite{He08a} two-dimensional assemblies
of tri-block copolymers built in three stages from a hierarchy of symmetrical
motifs,\cite{Groschel13a} and stacks of ordered discs that themselves have
self-assembled from rod-like virus particles.\cite{Barry10a}  However, it is not
a foregone conclusion that a target containing hierarchical structural motifs
necessarily assembles most reliably via a hierarchical
mechanism.\cite{Haxton13a}
\par
Inevitably, the choice of self-assembly strategy depends on the constraints in a
given case.  One-component assembly can only work for highly symmetric targets like
icosahedral capsids, and the strategy's frugality is achieved by considerable
sophistication of the building blocks themselves.  On the other hand, the exclusivity
of DNA brick interactions could represent a lavish overspecification in the case of
a simple target.  If working with building blocks that are less easily encoded than
DNA, it would be useful to know what are the simplest building blocks that could
self-assemble into a given target.
\par
The minimum amount of information required to specify a given target
structure depends both on the complexity of the target and the rules that govern the
assembly process itself.  In certain idealized cases, it is possible to quantify the
information in the minimal `kit' (structure plus rules).\cite{Ahnert10a,Johnston11a}
However, in more realistic situations that permit all the potential pitfalls of
kinetic assembly, it may be difficult to predict the minimal kit {\em a priori}.
In the high-information limit of fully addressable assembly, there are also limits
on self-assembly in terms of the yield of product\cite{Ke12a} and robustness of the
process with respect to the conditions.\cite{Reinhardt14a}  Despite the kinetic
nature of self-assembly, a sound understanding of the underlying thermodynamics
is always crucial, and the theory of stability in many-component mixtures must
be borne in mind.\cite{Sear04a,Jacobs13a}
\par
Self-assembly poses a number of challenges to simulation.  One ubiquitous difficulty
is spanning the full range of time- and length-scales involved.  Typically, particles
must diffuse through a medium to locate their binding partners and must then form
an aggregate that is capable of relaxing to the target structure.  If the mechanism
proceeds by nucleation, then target formation may formally be a rare event.
Furthermore, a self-assembling system is often highly inhomogeneous, starting from
a dilute solution of components and evolving towards a set of aggregates that are
locally quite compact and even more dilute than the original components when considered
as a species in their own right.  Under such conditions, and given complex building
blocks, straightforward molecular dynamics (MD) simulations are not necessarily the most
satisfactory way to proceed.
\par
Considerable insight can be gained from lattice-based
simulations,\cite{Reinhardt14a,Jacobs15a} and two-dimensional
models.\cite{Haxton13a} In such treatments, the interactions between particles
are usually implemented by a matrix of interspecies energies, rather than by an
explicit representation of the individual interaction sites that result in the
specificities encoded in the matrix.  For continuum models, it is sometimes
necessary to restrict the system to a single copy of the target structure (one
copy of each unique building block in the fully addressable
case\cite{Hormoz11a,Zeravcic14b}), thereby removing much of the potential
competition between partially completed structures that would be encountered
during self-assembly from a bulk phase.  An alternative to dynamical simulations
is a scheme based on Monte Carlo (MC) moves, provided that care is taken to move
aggregates\cite{whitelam_avoiding_2007,whitelam_role_2009} in such a way that
reproduces essential aspects of dynamics.  These methods have the advantage of
not requiring forces (or therefore derivatives of the potential), but can be
intricate to implement and are not guaranteed to produce physically realistic
diffusive motion.
\par
In this article, we initiate a comparison of self-assembly strategies for a
simple octameric target, starting from a minimal one-component approach,
proceeding to hierarchical multiple-step schemes and concluding with the fully
programmed limit of individually addressable sites.  The simulations are
performed on an idealized model (Section \ref{sec:Model}) of colloidal building
blocks that are cubic in shape and have a pattern of attractive sites explicitly
represented on their surfaces.  The particles are not confined to a lattice and
are free to rotate to any orientation.  The system contains multiple copies of
the building blocks required to construct the target structure, and the
simulations therefore incorporate the effects of competition between aggregates
at different stages of assembly.  Hence, although the model is highly
coarse-grained, it captures a number of important characteristics that occur in
real self-assembling systems, including some that are often neglected in
simulations.
\par
To follow the dynamics of self-assembly, we propose a hybrid Monte Carlo scheme
(Section \ref{sec:Monte Carlo Algorithm}) in which the internal relaxation of
clusters is handled separately from, but consistently with their diffusion.  We
also introduce a general scheme for identifying fragments of self-assembled structures
(Section \ref{sec:fragments}) to enable the histories of successfully
assembled targets to be traced and pathways of assembly to be elucidated.

\section{Model}
\label{sec:Model}

Our generic model for self-assembly consists of hard cubic particles, the faces
of which may be patterned with an arbitrary number of attractive patches. The
arrangement of patches on faces allows us to design pairs of faces with
complementary (or otherwise) interactions.  The interaction of particles through
patterned interfaces rather than via a single site on each particle captures an
important aspect of protein interactions, where binding involves an interface between
the surfaces of the proteins and determines their quaternary
structure.\cite{janin_protein-protein_2008}  Simplified representations of
protein interfaces as planar patterned surfaces have been used in some previous
theoretical work to investigate the distribution of overall interactions that
result when the abstracted surfaces approach.\cite{lukatsky_statistically_2006}
More recently, self-assembling single- and multi-component protein complexes
have successfully been computationally designed and experimentally realized,
based on detailed analysis of the interfaces between the
proteins.\cite{king_computational_2012,king_accurate_2014}  In the latter work,
the interactions were manipulated by altering the amino acid sequence of real
proteins to optimize the interfaces that would be required in the desired target.
\par
Although cubic particles may seem somewhat artificial in the context of proteins,
there are now many examples of synthetic routes to colloidal
cubes.\cite{sun_shape-controlled_2002,zhang_simple_2008,Rossi11a}  These
developments have stimulated both experimental and computational investigations
of the self-assembly of cubic
particles,\cite{Singh14a,vutukuri_experimental_2014,yang_controlled_2015}
and of their phase behavior.\cite{Smallenburg12a,Zhang11a}
While the experimental cubes do not as yet have patchy surfaces, we
note that, for spherical colloids, theoretical and computational work
on patchy particles\cite{Bianchi11a} has stimulated interesting and
fruitful experimental studies on spheres with directional
attraction.\cite{Pawar10a,Wang12b}
\green{It is also interesting to note that suspensions of macroscopic (centimeter scale)
cubes, which operate under a somewhat contrasting physical regime, have been
considered as building blocks for self-assembling modular robotic systems both
experimentally and in simulations.\cite{Tolley10a}}
\par
In order to detect overlap of the hard particle cores in the simulations,
we treat the cubes as oriented bounding boxes.\cite{gottschalk_obbtree:_1996}
The more general algorithms for detecting the overlap of two orthorhombic boxes are
then simplified to the case of cubes with edge $d$, following the same approach used for
simple hard cubes.\cite{john_phase_2005,Smallenburg12a}
\green{An alternative approach to modelling a polyhedron as a single object
is by fusing repulsive spheres into a rigid body.  Such models were
used in early simulations of colloidal cubes,\cite{John04a} although
the slight corrugations of the cube surfaces introduced some noticeable
artefacts.\cite{john_phase_2005}  Fused-sphere models with attractive spots
have also provided the basis for seminal work on the self assembly of virus
capsids.\cite{Rapaport99a,Rapaport10a}  The smooth faces of colloidal cubes
are an appealing blank canvas for experimenting with interaction patterns,
and we therefore adopt the hard cube model in this work.}
\par
\begin{figure}[tbp]
  \centering
  \includegraphics[width=\singlefigure]{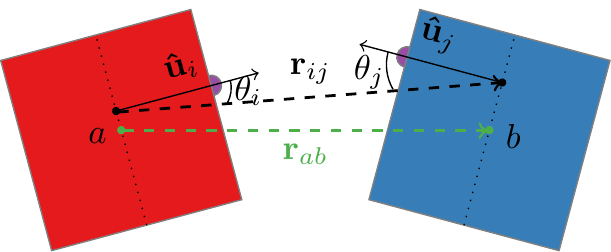}
  \caption{
Schematic representation of the interaction between patches $i$ and $j$ on two cubes
$a$ and $b$, showing the definition of the angles $\theta_i$ and $\theta_j$.
  }
  \label{fig:patches}
\end{figure}

\green{We implement patches on the faces of the hard cubes using a pairwise Morse
potential with an angular attenuation.}
The Morse potential between two sites corresponding to the patches $i$ and $j$
may be written
\begin{equation}
  V_{ij}^{\text{M}}(r_{ij}) = \varepsilon_{ij}
  \left[e^{-2\alpha(r_{ij} - d)} - 2 e^{-\alpha(r_{ij} - d)}\right],
  \label{eq:Morse}
\end{equation}
where $r_{ij}$ is the distance between the sites
and $\alpha$ is a parameter controlling the range of the potential.
We have chosen $\alpha = 6$, which gives a curvature at the minimum of the potential
that matches that of the Lennard-Jones potential.
$\varepsilon_{ij}$ is the strength of interaction between patches $i$
and $j$.  In general, this strength may vary between pairs of patches and so the
subscripts $ij$ are included on the function $V^{\rm M}_{ij}$ as well as on its argument
$r_{ij}$.  The Morse site representing a given patch is
embedded inside the cube at a depth $d/2$ from the surface with which the patch is
associated (\autoref{fig:patches}). Hence, the \green{repulsive core of the
patch potential coincides with the excluded core of the cube itself.  The attractive
tail of the Morse potential extends from the surface, with its minimum lying
at the point where the surface locations of the patches coincide.  This optimal
configuration therefore occurs when the cube faces are parallel and in contact.}
\par
We truncate the Morse potential at a distance $r_{ij}=2d$.  To avoid a discontinuity
at the cutoff, the potential is shifted by $V_{ij}^{\rm M}(2d)$ and
rescaled to recover a well depth of $\varepsilon_{ij}$.
The angular part of the potential takes the form of a Gaussian attenuation
\begin{equation}
  V^{\text{ang}} (\unit{r}_{ij}, \unit{u}_{i}, \unit{u}_{j}) = 
  \exp \left( - \frac{\theta_{i}^{2} + \theta_{j}^{2}}{2 \sigma^{2}}
  \right) ,
  \label{eq:ang}
\end{equation}
where $\unit{r}_{ij}$ is the unit vector pointing from patch $i$ to $j$.
$\theta_i=\cos^{-1}(\unit{r}_{ij}\cdot\unit{u}_i)$ and
$\theta_j=\cos^{-1}(\unit{r}_{ji}\cdot\unit{u}_j)$ are the angles between the
unit surface normals $\unit{u}_i$ and $\unit{u}_j$ of the patches and the
inter-patch vector (\autoref{fig:patches}).  The standard deviation $\sigma$ of
the Gaussian controls the width of the patches, by determining the rate at which
the potential decays with deviation from the ideal alignment.  In this work we
have fixed $\sigma$ at $0.2$.  \green{The definition of a patch as an embedded
  site, modulated by the angular attenuation of \autoref{eq:ang}, is directly
  analogous to that previously used in a Lennard-Jones-based model of patchy
  spheres.\cite{wilber_reversible_2007,Wilber09a,wilber_monodisperse_2009} One
  may envisage each patch as a conical region of attraction extending through
  the surface of the particle.  The directionality of the patch can be
  controlled by the Gaussian parameter $\sigma$ independently of the radial
range of the attraction.}

The overall form of the attractive potential between two patches $i$ and $j$
is therefore 
\begin{multline}
V^{\rm patch}_{ij}({\bf r}_{ij},\unit{u}_i,\unit{u}_j) = 
\left[\frac{V^{\rm M}_{ij}(r_{ij}) + V^{\rm M}_{ij}(2d)}
{\varepsilon_{ij} - V^{\rm M}_{ij}(2d)}\right] \times \\
\Theta(2d-r_{ij})
V^{\rm ang}(\unit{r}_{ij},\unit{u}_i,\unit{u}_j),
  \label{eq:potential}
\end{multline}
where ${\bf r}_{ij}=r_{ij}\unit{r}_{ij}$ and $\Theta$ is the Heaviside step function.
The total interaction between two cubes $a$ and $b$ is given by the sum of the 
interactions between the patches $i\in a$ and $j\in b$ on each of them:
\begin{multline}
V^{\rm cube}_{ab}({\bf r}_{ab}, {\mathbf\Omega}_a, {\mathbf\Omega}_b)
= \\ \sum_{i\in a} \sum_{j\in b} V^{\rm patch}_{ij}({\bf r}_{ij},\unit{u}_i,\unit{u}_j)
\Delta_{ij}(\unit{r}_{ij},{\mathbf\Omega}_a,{\mathbf\Omega}_b),
\end{multline}
where ${\mathbf\Omega}_a$ represents the orientation of cube $a$ and ${\bf
r}_{ab}$ is the vector position of the center of cube $b$ with respect to that
of cube $a$.  $\Delta_{ij}=1$ if the faces associated with patches $i$ and $j$
\green{are the} closest pair of most aligned faces of the two cubes, and
$\Delta_{ij}=0$ otherwise.  This restriction to interactions between only one
pair of faces at any instant effectively introduces a further truncation of the
interaction between pairs of patches, but the \green{strength of the interaction
  is negligible (typically less than $10^{-6}\varepsilon_{ij}$) at the point of
truncation} due to the angular attenuation of the potential as faces become less
aligned.
\par
The patchy cube model is versatile, allowing us to implement the different
self-assembly strategies outlined in Section \ref{sec:Introduction}.
The simplest system would consist of identical building blocks, each with
a small number of patches, all of which have the same interaction
energy.  Complexity can be introduced by more elaborate patterns of patches,
by multiple species of building blocks with different patch patterns, and by
specifying an alphabet of patch interaction strengths via the elements
$\varepsilon_{ij}$ of the matrix of pairwise well depths.

\section{Monte Carlo Algorithm}
\label{sec:Monte Carlo Algorithm}

\green{A full dynamical treatment of self-assembling building blocks would
  include the solvent or other medium in which the particles exist.  Such a
  level of detail is computationally expensive and can also be distracting.
  Effective dynamic schemes such as Langevin or Brownian dynamics capture some
  of the essential features of the solvent without representing it explicitly.
  Under certain conditions and with due care, it is also possible to reproduce
  dynamic-like behavior using Monte Carlo algorithms.  For example, such
  approaches have been useful in modelling relatively dense colloidal
  suspensions.\cite{Berthier07a,Romano11b} 
  \par 
  Monte Carlo is particularly appealing for models with a mixture of hard and
  continuous interactions,\cite{Ruzicka14b} such as the model detailed in
  \autoref{sec:Model}, since Monte Carlo does not require explicit forces and
torques and hence derivatives of the potential.} 
However, we envisage our self-assembling system being spatially highly
inhomogeneous because the overall suspension of building blocks is dilute, while
the assembling clusters themselves are locally dense.  The existence of
aggregates poses a problem for MC algorithms based on single-particle moves
because they only permit aggregates to move as a whole or to rearrange
internally by an energetically unfavorable and dynamically unrealistic shuffling
process in which attractive interactions are repeatedly broken and reformed.  It
is important to capture realistic diffusion in simulations of self-assembly,
since diffusion determines the rate at which building blocks and fragments of
structures encounter each other.  Equally, it is essential for aggregates to be
able to relax internally by collective motions.
\par
The symmetrized virtual move Monte Carlo (VMMC) algorithm of Whitelam and
Geissler\cite{whitelam_avoiding_2007,whitelam_erratum:_2008,whitelam_role_2009}
attempts to overcome the problems of single-move MC algorithms by constructing cluster
moves in response to the proposed trial move.  The algorithm recruits particles
to the cluster with a probability that depends on the difference in energy of the move
with and without each successive particle in the cluster.  VMMC thereby implicitly accounts
for \green{forces and torques generated by}
the potential without the need for derivatives to be calculated.
The VMMC algorithm produces natural collective motions
that allow proper internal rearrangement of aggregates.\cite{whitelam_avoiding_2007}
\green{The scheme has been deployed to obtain dynamic-like trajectories in various
strongly interacting systems.\cite{wilber_monodisperse_2009,Villar09a,Jacobs15b}  It
has also been shown to produce pathways and rates that are comparable with those from
Langevin dynamics in a coarse-grained model of DNA.\cite{Matek12a}
\par
As well as producing sensible cooperative motions within interacting groups of particles,
VMMC also improves the diffusion of small aggregates,}
compared to single-move Monte Carlo schemes.  However, diffusion still becomes
sluggish for large clusters in an uncontrolled way because the acceptance criterion for
trial moves is decreased by terms relating to the construction of the cluster in order to
satisfy detailed balance.
Here, we propose a general hybrid MC scheme that effects efficient internal relaxation
of clusters and realistic diffusion by treating these two types of motion separately but
consistently.

\subsection{Cluster diffusion}
\label{sec:diffusion}

At any instant, the self-assembling system may be unambiguously divided into
isolated clusters on the basis of interactions between the particles.  We define
any two cubes $a$ and $b$ for which $V^{\rm cube}_{ab}<0$ to be in the same
cluster.  A cluster is then defined by a network of such non-zero interactions.
\par
Half of the trial moves in our hybrid scheme are chosen to be diffusive steps of
isolated clusters.  For translational motion, these steps are implemented by
selecting a Gaussian-distributed random displacement along each Cartesian
direction.  For rotational motion, the same approach is taken to obtain a random
rotation vector through the cluster's center-of-mass.  To obey detailed balance,
neither type of diffusion move must produce a change in the system's
decomposition into clusters.  Hence, a move that causes energetic interaction
between previously isolated clusters must be rejected, as must any move that
produces a hard-core overlap between cubes.
\par
For a spherical object, the translational and rotational diffusion constants
vary with the particle's radius $R$ according to the Stokes--Einstein relations
\begin{subequations}
  \label{eq:brownian}
\begin{align}
  D_{\mathrm{t}} &= \frac{ \kb T }{ 6 \pi \eta R } , \quad\text{and}\\
  D_{\mathrm{r}} &= \frac{ \kb T }{ 8 \pi \eta R^{3} } ,
\end{align}
\end{subequations}
where $\kb$ is Boltzmann's constant, $T$ is the temperature and $\eta$ is the
viscosity of the medium.  Our main concern in simulations of self-assembly is
that, between collisions with other aggregates, clusters of different numbers of
particles should diffuse at physically reasonable relative rates and that the
translational and rotational diffusion constants for a given cluster should be
related by
\begin{equation}
  \frac{ D_{\mathrm{t}} }{ D_{\mathrm{r}} } =
  \frac{ 4 }{ 3 } R^{2},
  \label{eq:brownian_ratio}
\end{equation}
in accordance with \autoref{eq:brownian}.  For simplicity, we approximate
clusters of $n$ particles in our simulations as spheres of radius proportional
to \green{$n^{1/3}$}.  Since the diffusion constant of a random walk is
proportional to the square of the mean step size, the mean translation and
rotation steps should then vary with cluster size according to
\begin{subequations}
\begin{align}
  \delta_{\mathrm{t}} (n) &\propto n^{-1/6} \quad\text{and}
  \\
  \delta_{\mathrm{r}} (n) &\propto \sqrt{ \frac{3}{4}}n^{-1/2}
\end{align}
\end{subequations}
respectively.
\par
In practice, diffusive steps are implemented by selecting a particle at random,
determining the other particles that belong to the same cluster, and then performing
the trial move of the cluster.  However, by this method, the probability of
choosing a particular cluster is proportional to the number of particles it
contains, since any of its members could have been the particle initially selected.
In diffusive motion, the mean square distance travelled is proportional to the number
of steps.  We may therefore cancel the effect of choosing a cluster with probability
proportional to $n$ by reducing the step size by a factor $n^{1/2}$.  Hence, to
produce the correct relative mean-square displacement of aggregates in a given
portion of simulation trajectory, our final choice of step sizes is
\begin{subequations}
  \label{eq:delta_scaling}
\begin{align}
  \delta_{\mathrm{t}} (n) &\propto n^{-2/3} \quad\text{and}
  \label{eq:delta_scaling1} \\
  \delta_{\mathrm{r}} (n) &\propto \sqrt{ \frac{3}{4}}n^{-1}.
  \label{eq:delta_scaling2}
\end{align}
\end{subequations}
\par
Given \autoref{eq:delta_scaling}, all diffusion step-size parameters are fixed once
the diffusion of an isolated monomer has been chosen.  The monomer step size will be
determined in Section \ref{sec:relaxation}.
\green{The scalings in \autoref{eq:delta_scaling} therefore remove any arbitrariness
in the choice of step size for different clusters on the basis of the Stokes--Einstein
relations, subject to the approximation of a compact shape.  For our purposes, this
controlled decrease in diffusion rate with cluster size is sufficient.  However, we
note that it would be possible to refine the approach by calculating the largest
diameter of a given aggregate in the direction of travel if desired.\cite{whitelam_avoiding_2007}}

\subsection{Internal relaxation}
\label{sec:relaxation}

The remaining half of the MC steps in our hybrid scheme are
collective moves for internal relaxation of clusters, performed using
symmetrized VMMC as described in Ref.~\citenum{whitelam_role_2009}.
The algorithm builds a subcluster that is appropriate for the proposed
translational or rotational move.  To avoid interference with the bulk
diffusion of isolated clusters described in Section \ref{sec:diffusion},
a VMMC move must be rejected if it recruits all the particles in an
isolated cluster and proposes a move that leaves the same cluster isolated.
This limitation permits VMMC moves to join formerly isolated clusters,
to separate an isolated cluster into two clusters, and to effect internal
relaxations of a cluster by motion of a subcluster, but not merely to
move an isolated cluster while keeping it isolated.  This list of
operations is exactly complementary to those covered by the diffusive
moves described in Section \ref{sec:diffusion}.
\par
In VMMC, the building of a subcluster starts with the displacement of a randomly
chosen seed particle.  We found the behavior of the VMMC algorithm to depend on
the size of steps attempted, with small steps tending to move only single
particles, and large steps generally attempting to move all interacting
particles. In order that the VMMC algorithm is able to efficiently relax, form
and break clusters, we select a step size parameter such that the VMMC moves
attempt to displace aggregates of intermediate size with a reasonable frequency.
The corresponding mean trial displacement is then transferred to the diffusion
steps described in Section \ref{sec:diffusion} for the displacement of isolated
monomers, thereby providing a smooth handover between the two parts of the
algorithm at the single-particle level.
\par
The motion of a subcluster in a VMMC move can alter the center-of-mass
of the isolated cluster to which it belongs.  Hence, internal relaxation
moves do make a small contribution to the diffusive motion of isolated
clusters.  However, we have found this effect to be small enough not
appreciably to disrupt the desired Stokes--Einstein-like relative diffusion of
different sized clusters.
\par
\begin{figure}[tbp]
  \centering
  \includegraphics[width=\singlefigure]{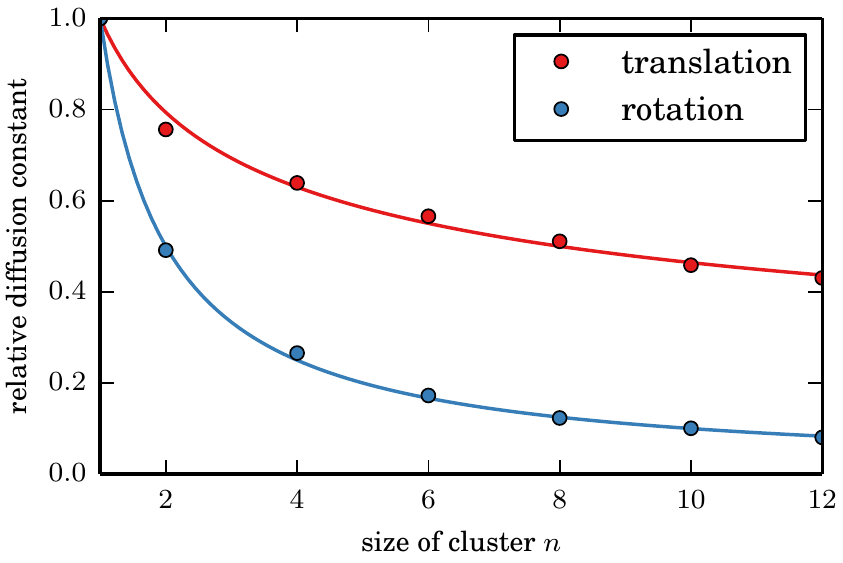}
  \caption{
Diffusion constants in the low density limit for clusters of up to
12 particles, relative to the value for a monomer.  Circles are from
simulations with the hybrid MC algorithm and lines are the Stokes--Einstein
equations for spherical particles.
  }
  \label{fig:diffusion}
\end{figure}

The performance of the hybrid MC algorithm
with respect to diffusion is demonstrated in Fig.~\ref{fig:diffusion},
which shows diffusion constants, relative to those for a single particle,
as a function of cluster size for clusters that are sufficiently tightly
bound to retain their integrity over a long period.  The diffusion constant
of each cluster is obtained by simulating a single cluster of the required
size.  The mean squared displacement $\langle\Delta r^2\rangle$ is then
related to time by $\langle\Delta r^2\rangle = 6D_{\rm t}t$, where time
$t$ is measured in MC steps for now.  Rotational diffusion is estimated
analogously, representing angular displacements in vector form and summing
them to obtain the overall angular distance travelled.\cite{Romano11b}

\subsection{Time scale}

The diffusion constant of an isolated particle provides a reference for
interpreting the number of MC steps as a time scale.  This diffusion constant
is in turn fixed by the step sizes in the VMMC part of the simulation
algorithm, as described in Section \ref{sec:relaxation}.  We have found that
selecting Gaussian-distributed components of displacement vectors with a standard
deviation of $0.2d$ provides a satisfactory acceptance rate for VMMC steps
without the need for adjustment over a wide range of temperatures.
\par
If MC steps were used as the unit of time, the MC algorithm as described
would produce temperature-independent diffusion constants of aggregates
in the low density limit.
However, the diffusion constants should vary with temperature according
to \autoref{eq:brownian}.  We therefore use the Stokes--Einstein equations as a basis
for mapping the relative time scales of simulations performed
at different temperatures
onto the number of MC steps.  For example, if the temperature is doubled,
the time notionally associated with an MC step should be halved so that the
diffusion constant with respect to this scaled time is effectively increased
by a factor of 2.  We therefore arrive at the mapping
\begin{equation}
t^* = n_{\rm cycles}/T^*
\label{eq:tstar}
\end{equation}
from the number $n_{\rm cycles}$ of MC cycles to reduced time $t^*$ using a
reduced temperature $T^*$ that will be defined in Section \ref{sec:results}.

\section{Fragment detection}
\label{sec:fragments}

In order to quantify the progress of assembly in a simulation, we will need
to detect successfully completed target structures, as well as plausible
intermediates and fragments.  All such clusters will deviate from idealized
geometries due to thermal fluctuations, making it necessary to introduce
some tolerance in matching instantaneous configurations to a library of
structures being tracked.  The method for identifying fragments must be
computationally efficient, since it will be applied repeatedly during
the simulations.  In particular, the algorithm must be able to cope with
the arbitrary position and orientation of the fragment, and be invariant to 
permutation of indices of identical particles.
\par
A class of metrics that satisfy many of the desirable properties of fragment
detection algorithms is based on matrices of a pairwise property such as
distance or energy.  The eigenvalues of such matrices are unaffected by bulk
translation, rotation and particle permutations, thereby providing a
`fingerprint' by which structures can be recognized.\cite{sadeghi_metrics_2013}
Nevertheless, a tolerance for comparison of the eigenvalues must be chosen, and
the most appropriate tolerance may depend on the size of the structure and the
number of components in its fingerprint.  Furthermore, an eigenvalue represents
a collective property; therefore, a deviation from a reference value only as a
delocalized structural interpretation.  Instead, we have devised a scheme for
fragment recognition that is based on pairwise links between particles.  The
parameters in the method have direct geometric interpretations and can be fixed
for fragments of all sizes.
\par
The first step is to identify aggregates that are potential candidates for
recognition as a structural fragment.  Here, we use a similar definition of aggregates
to that used for isolated clusters in diffusion MC moves (Section
\ref{sec:diffusion}), i.e., that cubes $a$ and $b$ belong to the same cluster
if they are interacting, $V^{\rm cube}_{ab}<0$.  However, for the purposes of
defining an aggregate, we also impose the requirement that $a$ and $b$ are
closer than $r_{ab}=1.207d$, which is the shortest distance between the centers of
two cubes when their orientations differ by $45^\circ$.
This additional criterion admits particles that may be interfering with a structural
fragment and should not be ignored.  However, the criterion does trim very
loosely associated particles from the aggregate to avoid the possibility that
a fragment will not be recognized on the basis of other particles that happen
to be in the vicinity.  In well defined fragments that are clear of other aggregates,
all interparticle distances within the fragment
lie decisively below this initial criterion,
making the assignment of particles to the aggregate unambiguous in most cases.
A quick test can now be performed on the aggregate to see whether its
size and (in cases where more than one type of building block is in use) its composition
match those of recognized fragments\green{; any aggregates not recognized at
this stage need not undergo structural analysis at all.}
\par
\begin{figure}[tbp]
  \centering
  \includegraphics[width=\singlefigure]{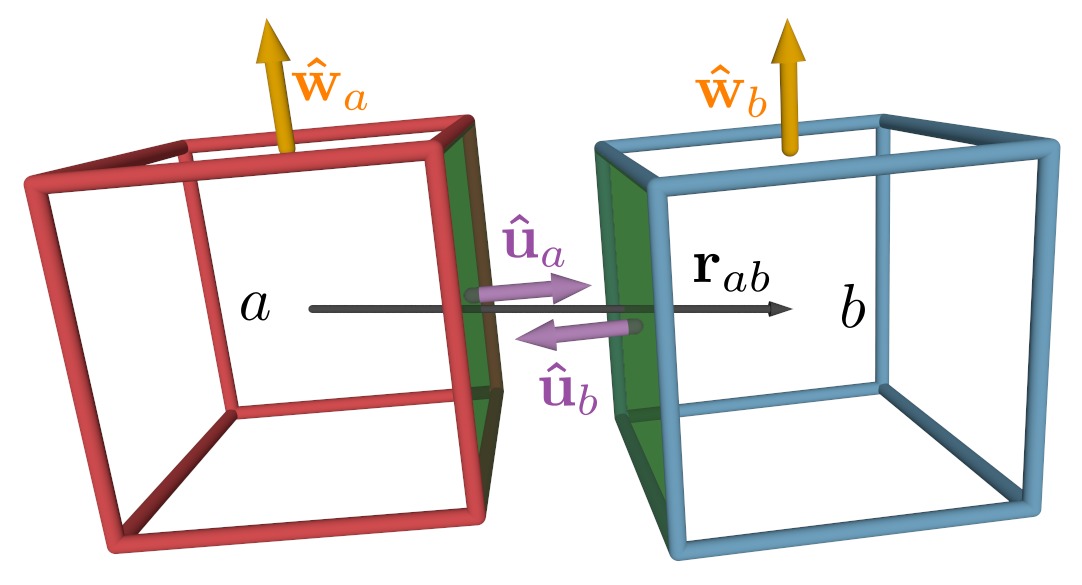}
  \caption{
Testing of links between particles in a fragment involves the unit normals
$\unit{u}_a$ and $\unit{u}_b$ to the faces that have come together, and the
auxiliary vectors $\unit{w}_a$ and $\unit{w}_b$ associated with the linked
faces.
  }
  \label{fig:link}
\end{figure}

The second step examines the aggregate more closely by enumerating the links
between particles.  A link between two neighboring particles $a$ and $b$
is characterized by the species of the two particles and by the labels of
the particular two faces that are closest.  To qualify as a link, three criteria
must be met: (i) the centers of the building blocks must lie closer than a
more stringent distance of $r_{ab}=1.140d$, (ii) the faces that approach each
other must be sufficiently aligned, and (iii) the building blocks must lie at
the correct relative orientation.  The criteria for alignment in (ii)
are $\unit{r}_{ab}\cdot\unit{u}_a>0.8$ and $\unit{r}_{ba}\cdot\unit{u}_b>0.8$,
where $\unit{u}_a$ is the outward unit normal of the relevant face of particle $a$
(see \autoref{fig:link}).
The criterion for relative orientation in (iii) is $\unit{w}_a\cdot\unit{w}_b>0.95$, where
$\unit{w}_a$ is an auxiliary unit vector attached to one of the faces of cube $a$
adjacent to the face defining the contact, and $\unit{w}_b$ is its counterpart
on cube $b$, chosen such that $\unit{w}_a$ and $\unit{w}_b$ are parallel in
the ideal fragment geometry.  As shown in \autoref{fig:link},
the auxiliary vectors effectively define a torsional
angle upon which a tolerance is placed.  We emphasize that the auxiliary vector
is associated with the link on a particular face.  A given building block has an
auxiliary vector associated with each of the faces through which it may form a link.
The tolerances in the face alignment
and torsional criteria allow for thermal fluctuations.  Particles that are not
part of a well structured fragment typically fail these criteria by a wide margin.

A given fragment that is to be tracked in the simulation
is specified by the list of links that define its geometry.
Recognition of a fragment that arises in a simulation is simply a matter of
matching a sorted list of links (each defined by the two particle species and
two face labels) against the lists of fragments that are being tracked.
\par
The procedure described here is geometrically intuitive for the cubic building
blocks used in the present work.  However, the approach can readily be applied in
other models of self-assembly by attaching appropriate alignment vectors
$\unit{u}$ and auxiliary orientation vectors $\unit{w}$ to building blocks.

\section{Results}
\label{sec:results}

We will contrast self-assembly strategies for a simple octamer
target, consisting of eight building blocks joined into a $2\times2\times2$
cube.  The octahedral symmetry of this object can be exploited to
deploy the full spectrum of strategies from minimal one-component cases
to fully addressable eight-component mixtures.
\par
The simulations are performed with 64 monomers (always in the correct quantities
to make eight copies of the target possible) at a packing fraction of 0.05.
Simulations are initiated from a very high temperature run, where the
equilibrium state is a `gas' of monomers, and quenched instantaneously to
the desired temperature at the start of the assembly run.  To obtain
statistics on the progress of assembly as a function of time, an ensemble of
50 replicas starting from different disordered configurations are
typically averaged under any given
set of conditions.  To facilitate comparison between different
building block designs, the thermal energy is always referenced to the energy
$E_{\rm target}$ of the optimized target structure, thereby defining a
reduced temperature $T^*=k_{\rm B}T/|E_{\rm target}|$.  This, in turn,
defines the relative time scale of the MC simulations through \autoref{eq:tstar}.

\subsection{Sequential assembly}
\label{sec:seq}

\begin{figure}[tbp]
  \centering
  \includegraphics[width=\singlefigure]{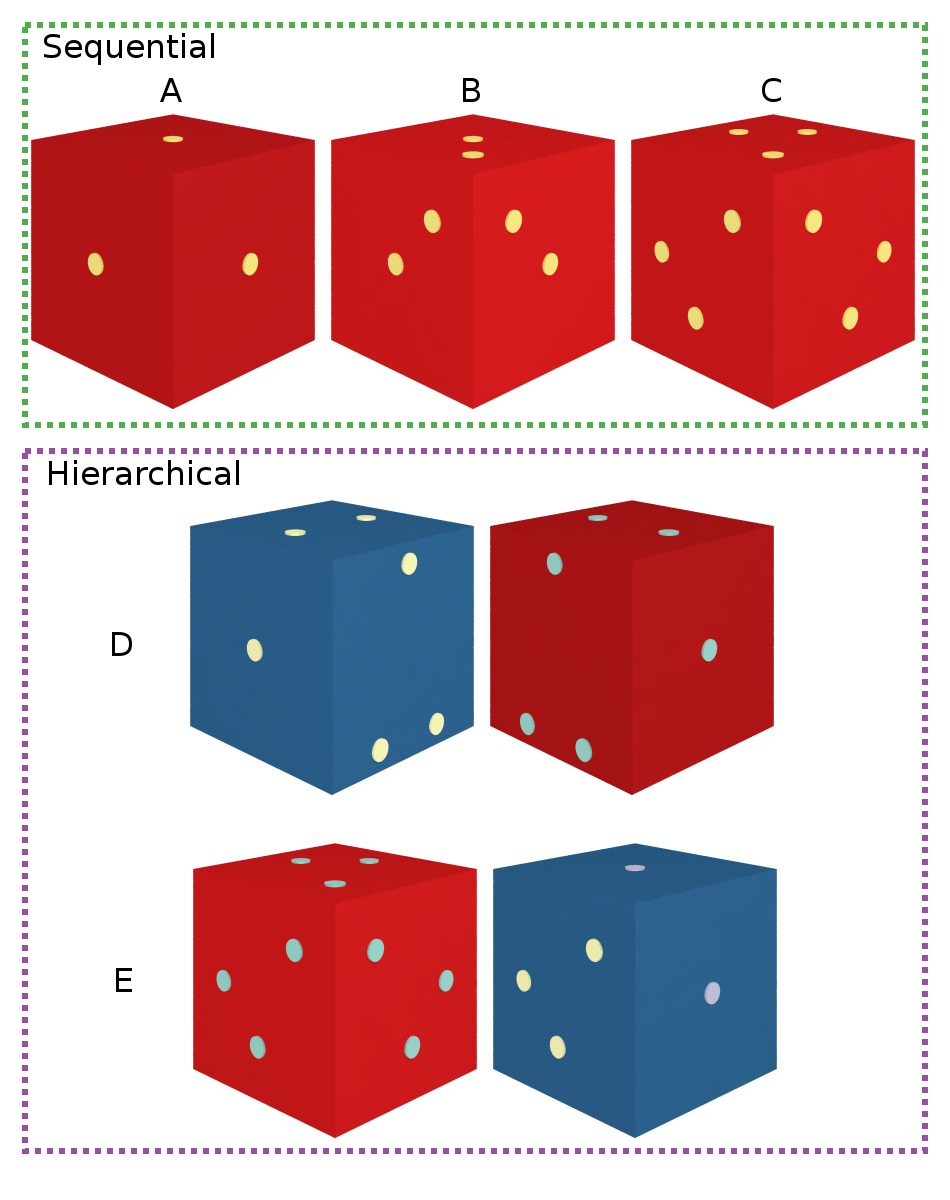}
  \caption{
    Designs of patchy particles used in this work. The sequential A, B and C
    models self-assemble from systems of only one component and were designed to
    test the minimal requirements for self-assembly.  Both hierarchical models,
    D and E, consist of two particle types and use different numbers of patches
    per face, in conjunction with interaction alphabets, to assemble in multiple
    steps.
  }
  \label{fig:models}
\end{figure}

The octamer target can be assembled from minimal models consisting of a single
species of building block and a single type of attractive patch if the patterned
building blocks have $C_{3v}$ symmetry.  The diagonal line of symmetry on each
face then ensures that the patches on facing cubes match as the target is
assembled.  Three designs, A--C, with this symmetry constraint are illustrated
in the top row of \autoref{fig:models}.  In these designs, all pairs of patches
interact with the same strength $\varepsilon_{ij}=\varepsilon$.  We simulate
each model at a range of temperatures to identify both the optimal temperature
for assembly and the reliability of assembly with deviation from the optimum.
We expect the target to assemble by the sequential addition of building blocks
to a growing structure.

\begin{figure*}
  \centering
  \includegraphics[width=\doublefigure]{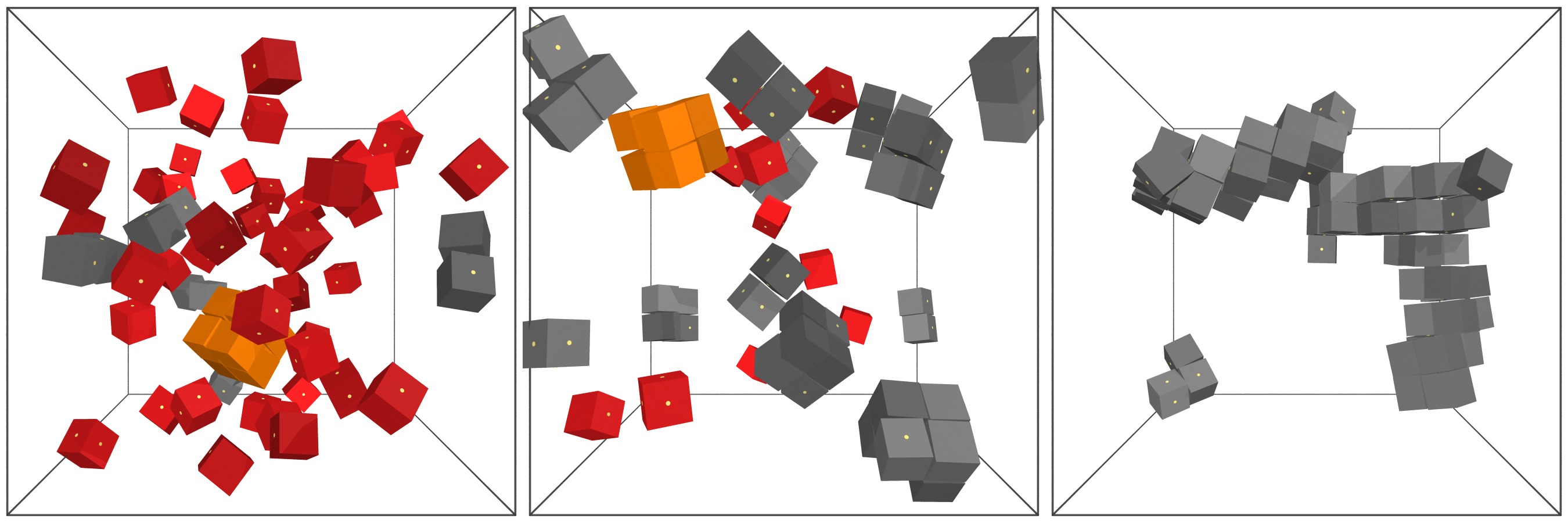}
  \caption{
    Snapshots from simulations of model A at $\Tstar = 0.07$, $0.06$ and
    $0.04$ (left to right). In the left snapshot we see a single correctly
    assembled octamer (orange), with most particles existing as monomers (red).
    At a lower temperature (center), we see isolated aggregates (gray) with
    incorrect orientation of building blocks, and at very low temperatures
    (right) the particles form large, extended aggregates.
  }
  \label{fig:seq_simple}
\end{figure*}

In model A, which has one patch in the center of each of three adjacent faces,
12 pairs of patches come into contact in the perfect target structure,
giving an energy of $E_{\rm target}=-12\varepsilon$ and defining a reduced
temperature $T^*=k_{\rm B}T/12\varepsilon$.  Although the target structure
optimizes the energy of the system, a vast number of disorganized networks
also allow all (or most) pairwise interactions between patches to be satisfied.
Hence, model A proves to be a poor design, with very few correct targets observed
(\autoref{fig:seq_simple}). The temperature window in which
the target cluster is seen is also very narrow, around $0.06\lesssim\Tstar\lesssim0.07$.
Above this range the system exists as a monomer fluid, and below it
only large aggregates are seen.
At $\Tstar = 0.06$ clusters resembling the target, or fragments of the target
arise.  However, particles within these clusters are often oriented
incorrectly, leaving dangling patches. These aggregates will not form the target
structure without breaking up first, and may instead join with other aggregates to
form large amorphous structures, as seen at lower temperatures
(\autoref{fig:seq_simple}).

Building blocks in model B (\autoref{fig:models}, top middle) have additional
detail on the patterned faces whilst retaining the $C_{3v}$ symmetry of model A.
The second patch on each patterned face introduces an orientational preference,
leaving two configurations of a dimer that align both pairs of patches.  Only
one of these orientations leaves all faces of both particles lying in directions
that are compatible with the target structure.  Although the incorrect
orientation has the same dimer energy, we have tried to discourage it from
forming by offsetting the two patches from the center of the building blocks'
faces.  Incorrectly bound dimers therefore have staggered faces, which helps to
suppress further growth of the faulty structure on steric grounds.
\par
The additional patch in model B amounts to the explicit implementation of a
torsional potential.  Torsion has been shown to play an important role in self
assembly in less detailed models where the torsional effect is included in the
potential directly as a function of the dihedral angle between the building
blocks.\cite{wilber_monodisperse_2009} As expected, therefore, model B assembles
much more reliably than model A.  The additional information encoded in the
particles (doubling the number of patches compared to the ineffective model A)
has produced what is probably the minimal viable design for the octamer target.
\par
\begin{figure}[tbp]
  \centering
  \includegraphics[width=\singlefigure]{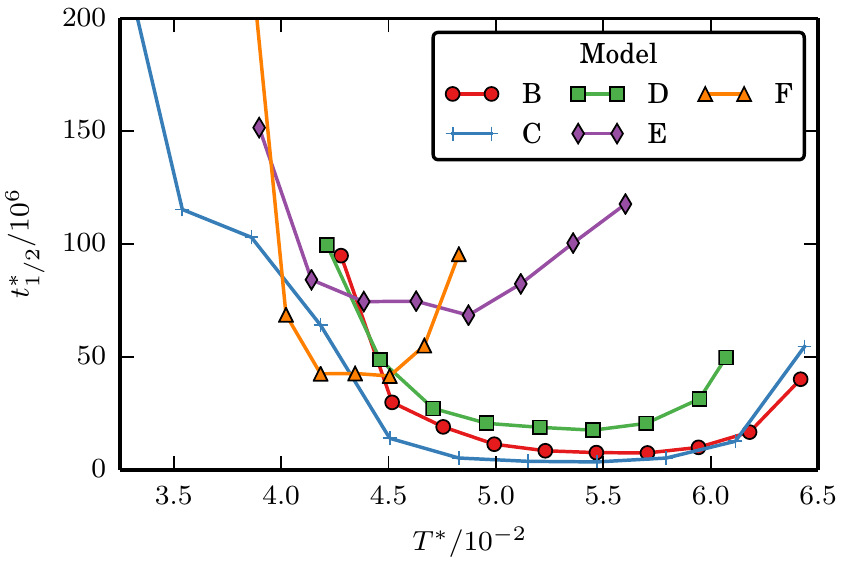}
  \caption{
    The time $t^*_{1/2}$ taken to achieve $50\%$ yield of the octamer target as a
    function of reduced temperature for models B--E.
  }
  \label{fig:threshold}
\end{figure}

To quantify the efficiency of assembly, we measure the time $t^*_{1/2}$ taken
for a given model to incorporate 50\% of its monomers into completed target
structures.  This time is shown as a function of temperature for model B by the
red circles in \autoref{fig:threshold}.  The optimal temperature for
self-assembly is around $T^*=0.055$, but there is a broad range around this
value where assembly proceeds at a comparable rate.  The limiting factor at high
temperature is the decreasing thermodynamic stability of the target with respect
to monomers.  At low temperature, assembly suffers from the longer time required
for errors to be corrected, amounting to a kinetic trap.

Model C for sequential assembly (\autoref{fig:models}, top right) contains a
further orientational constraint, with each patterned face decorated by an
isosceles triangle of patches.  An equilateral triangle is deliberately avoided,
since it would result in three degenerate bindings of a dimer, only one of which
is correct.  Model C therefore reduces the weakness of model B by leaving only
one relative orientation in which all patches can be simultaneously satisfied in
a dimer of building blocks.  Weakly bound incorrect structures are still
possible by alignment of two pairs of patches, but such structures are both
energetically less stable and geometrically less compatible with the target.
\par
The blue trace in \autoref{fig:threshold} shows that model C has a slightly
lower optimal assembly temperature than model B and that it assembles more
quickly and over a wider range of temperatures.  The extra information added to
the faces to make model C has effectively been used for negative
design,\cite{Richardson89a} destabilizing incorrectly bound building blocks.

The three cases A--C establish that robust assembly of our highly symmetrical
target structure can be obtained in a one-component system with a one-letter
alphabet of interaction types.  For successful assembly,
interactions between building blocks must
place a preference on relative orientation.  Two patches per face (model B)
are enough to implement this bias at a minimal level but performance can be
further improved by adding a third patch (model C).  The
third patch refines the effective torsional potential
between building blocks, biasing the binding towards the desired orientations.
\par
We note that the optimal temperature for assembly in the one-component
sequentially-assembled models is quite high, with a thermal energy $k_{\rm B}T$
that corresponds to about 70\% of the maximum interaction energy between two building blocks.
At such temperatures, connections between individual building blocks are transient,
allowing the system to escape from kinetic traps.  Furthermore, partially completed structures
can disintegrate, allowing their building blocks to be incorporated elsewhere.

\subsection{Hierarchical assembly}
\label{sec:hier}

An alternative to building up the target by sequential addition of monomers
is for sections of the target to be assembled independently and then joined.
In this section, we test two schemes that promote such hierarchical pathways.
Both schemes deploy a binary mixture of particle designs.

\begin{figure}[tbp]
  \centering
  \includegraphics[width=\singlefigure]{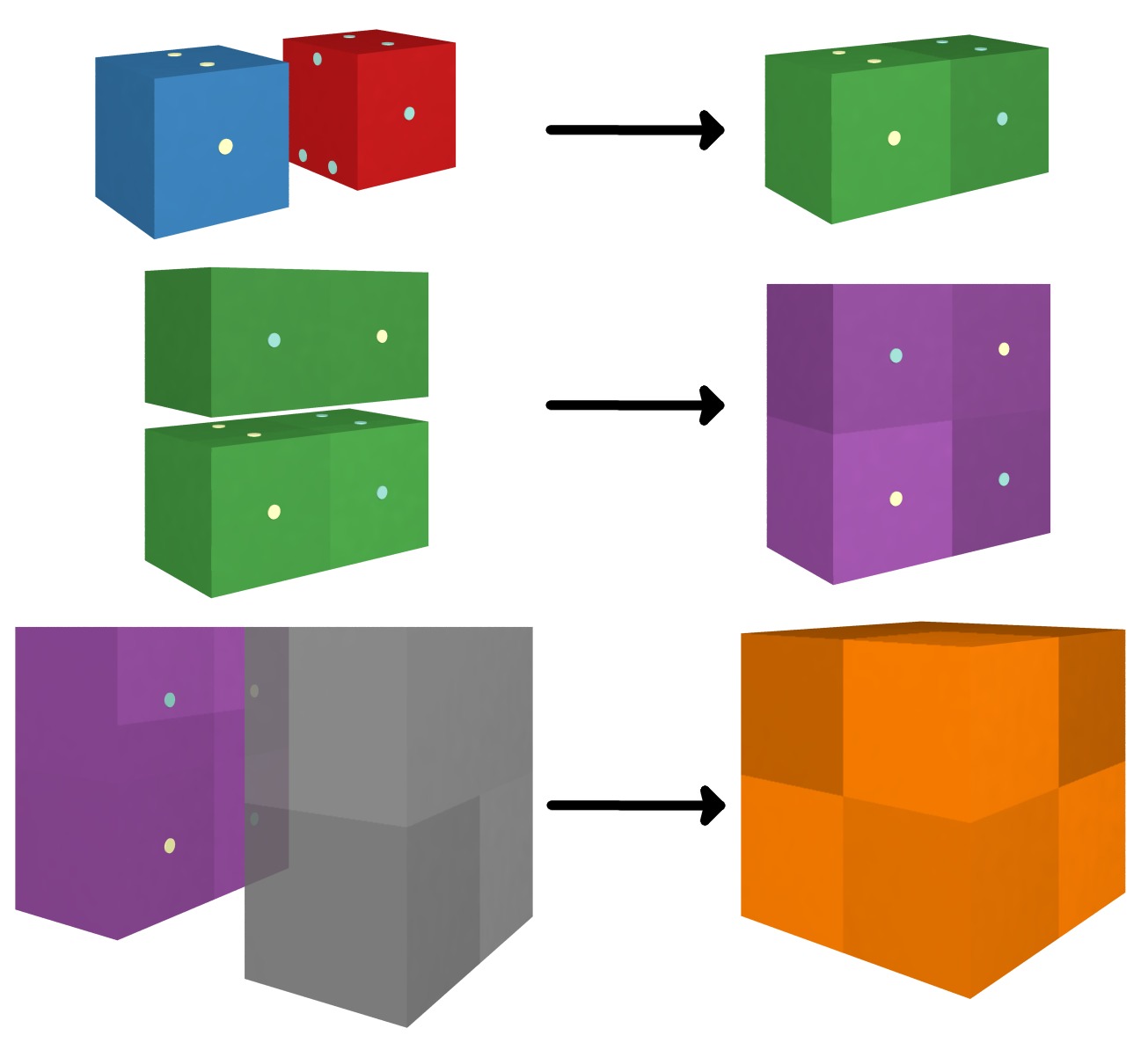}
  \caption{
    The intended hierarchical assembly pathway of model D. A
    pair of one D1 (red) monomer and one D2 (blue) monomer bind through
    their three-patch faces to form a dimer (green).  Two dimers bind through
    their four-patch faces to form a tetramer (purple).  A pair of tetramers
    then bind through their only patterned faces to form the octamer target
    (orange).
  }
  \label{fig:hier_assembly}
\end{figure}

In model D, the two species (D1 and D2) are mirror images of each other
(\autoref{fig:models}, middle row).
Patches on opposite particle types attract with equal strength
$\varepsilon_{ij}=\varepsilon$, but patches on identical particles have
no interaction with each other, $\varepsilon_{ij}=0$, making a two-letter
alphabet with an anti-diagonal interaction matrix.
Despite the single strength of interacting patches, corresponding pairs of
faces on the two cube types can be given different interaction energies by
having different numbers of patches.  To encourage assembly events to occur
in a particular order, we decorate one face with three patches, one with
two and one with a single patch.  The expectation is that a three-step
assembly will take place (\autoref{fig:hier_assembly})
in which a D1--D2 dimer forms first, by
binding on the three-patch faces.  The resulting intermediate has one
four-patch face and one two-patch face, promoting formation of a tetramer
via the four-patch faces.  Finally, two tetramers then align their four pairs
of patches to make the target.

\begin{figure}[tbp]
  \centering
  \includegraphics[width=\singlefigure]{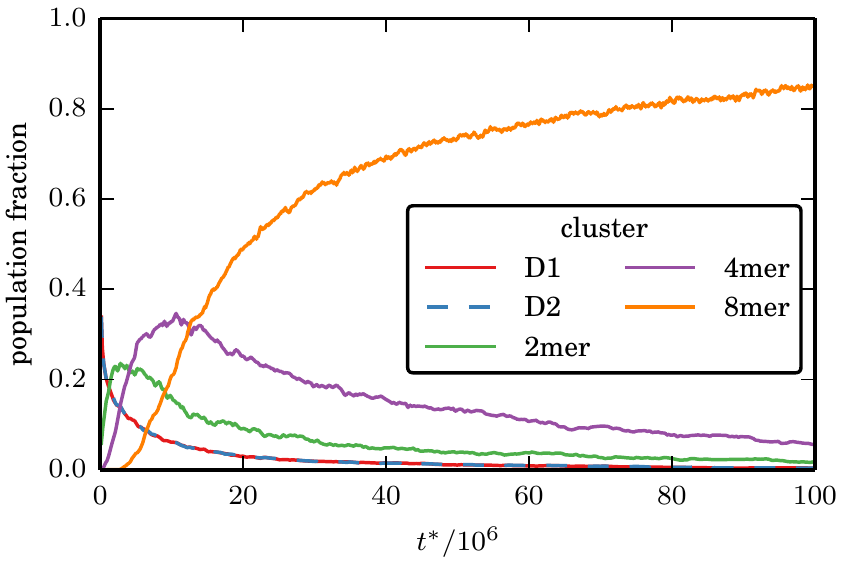}
  \caption{
    The fraction of building blocks in the hierarchical model D existing as
    monomers, as part of a complete target, or in correct fragments of
    the target, as a function of time averaged over 100 simulations at $\Tstar = 0.05$.
    The population of 5-, 6- and 7-member clusters is too low to be visible
    on the scale of this plot, and the lines corresponding to the two monomer types
    coincide almost exactly.
  }
  \label{fig:hier_assembly_5}
\end{figure}

Evidence that assembly takes place by the intended hierarchical route
is provided by \autoref{fig:hier_assembly_5}, which shows a rapid decay of the
population of isolated monomers as they become incorporated into larger structures,
and successive peaks in the populations of dimers and tetramers before the
population of completed target builds up.  Crucially, the population of
trimers is never high, implying that very few tetramers form by sequential
addition of monomers to a dimer.  Similarly, the populations
of \mbox{5-,} 6- and 7-member clusters are too low to be visible
in \autoref{fig:hier_assembly_5}.
\par
\begin{figure}[tbp]
  \centering
  \includegraphics[width=\singlefigure]{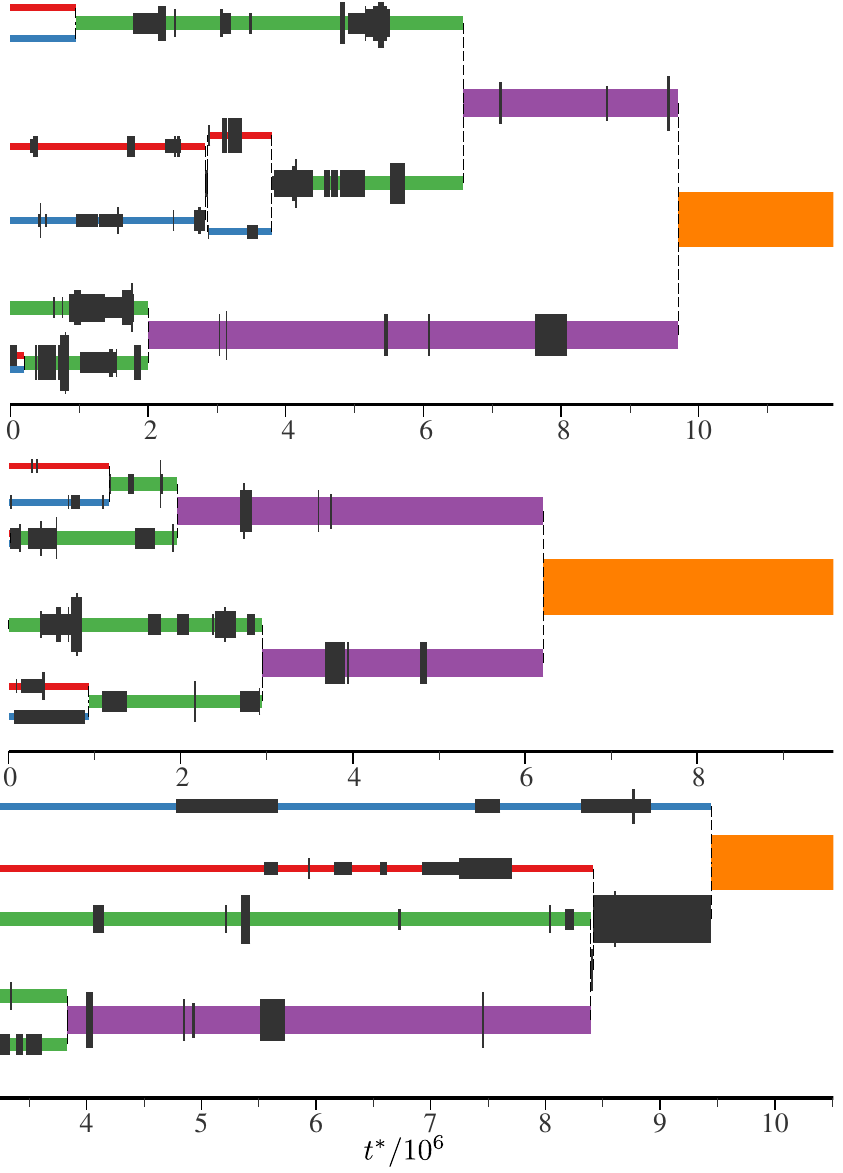}
  \caption{
    Extracts from the history of target clusters in the hierarchical model D,
    taken from one simulation at $\Tstar = 0.05$.  Colored lines correspond
    to the scheme in \autoref{fig:hier_assembly}.
    The first two histories are
    examples of clear-cut hierarchical assembly, while the third history shows
    an alternative route where a cluster of seven particles is formed and a
    monomer then joins to complete the octamer.
  }
  \label{fig:hier_history}
\end{figure}

Direct information on individual trajectories can be obtained by examining
the history of successfully completed targets.
\autoref{fig:hier_history} gives three examples of cluster histories for
model D.  In these diagrams the horizontal axis represents time since the
start of the assembly simulation.  The horizontal lines represent clusters,
with the thickness proportional to the number of particles in the cluster and
the color identifying recognized fragments using the same convention as
\autoref{fig:hier_assembly_5}.  Black segments indicate unrecognized aggregates.
Thin dashed lines indicate the joining or fission of clusters.  \green{A cluster is
only included in the diagram if one of its particles appears in the final cluster.
Hence, tracing the diagram from right to left reveals the history of how particles
came to be incorporated in the product.  Other particles that interact only transiently
with those that are finally incorporated appear as a temporary thickening of the
corresponding line.}
\par
The first two histories in \autoref{fig:hier_history} show decisively hierarchical
paths to the target, with lines representing monomers, dimers and tetramers joining in
turn.  Brief black segments on the colored lines indicate that
additional building blocks temporarily attach to the developing structure,
but these excursions always revert to the underlying
fragment in the examples shown.  The third history in \autoref{fig:hier_history}
shows that sequential paths are not ruled out in model D; in this example
at $t^*\approx8.5$, a tetramer, a dimer and a D2 monomer combine to make a heptamer,
with the final D1 monomer being added some time later.

\begin{figure}[tbp]
  \centering
  \includegraphics[width=\singlefigure]{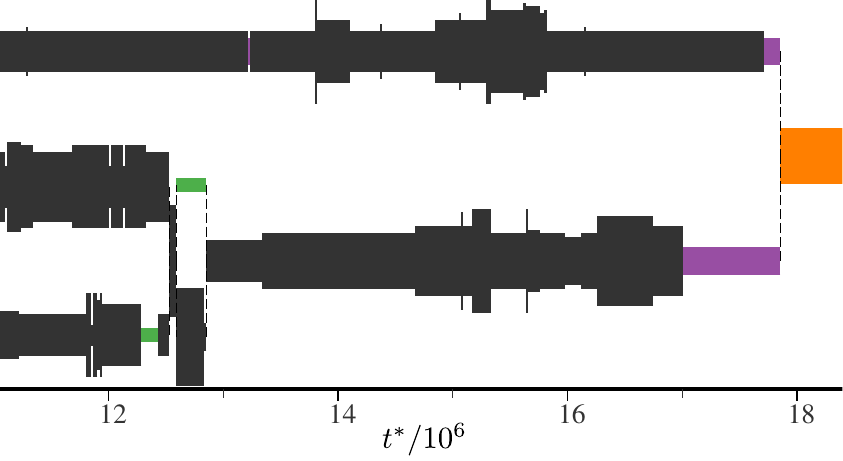}
  \caption{
    An extract from the history of an octamer forming at $\Tstar = 0.04$
    in model D.  Two correct tetramer fragments break apart from larger,
    unrecognized aggregates and then bind to form the octamer target.
  }
  \label{fig:hier_budding}
\end{figure}

Even for a given model, the mechanism of assembly can change with conditions.
At the low temperature of $\Tstar = 0.04$ we see a late, but steady
production of the target cluster in model D.  However, at this temperature
the assembly is not hierarchical.  Large, disordered aggregates form rapidly
and, through internal rearrangement,
correct subclusters may be released from a larger aggregate, and then proceed to
form the target structure.  In the example shown in \autoref{fig:hier_budding},
two tetramers emerge from separate larger aggregates and then combine directly
to make the target.  This path to assembly is similar to the budding
mechanism identified in a one-component system of patchy
spheres.\cite{wilber_reversible_2007}

The green trace in \autoref{fig:threshold} shows that model D is quite
robust with respect to changes in temperature, but the simpler sequential models
B and C both do better in terms of speed and temperature range.  The strategy of
assembling sub-components of the target simultaneously, which comes at the
expense of more complex building blocks, does not seem to have
provided model D with an advantage.\cite{Haxton13a}
\par
\begin{figure}[tbp]
  \centering
  \includegraphics[width=\singlefigure]{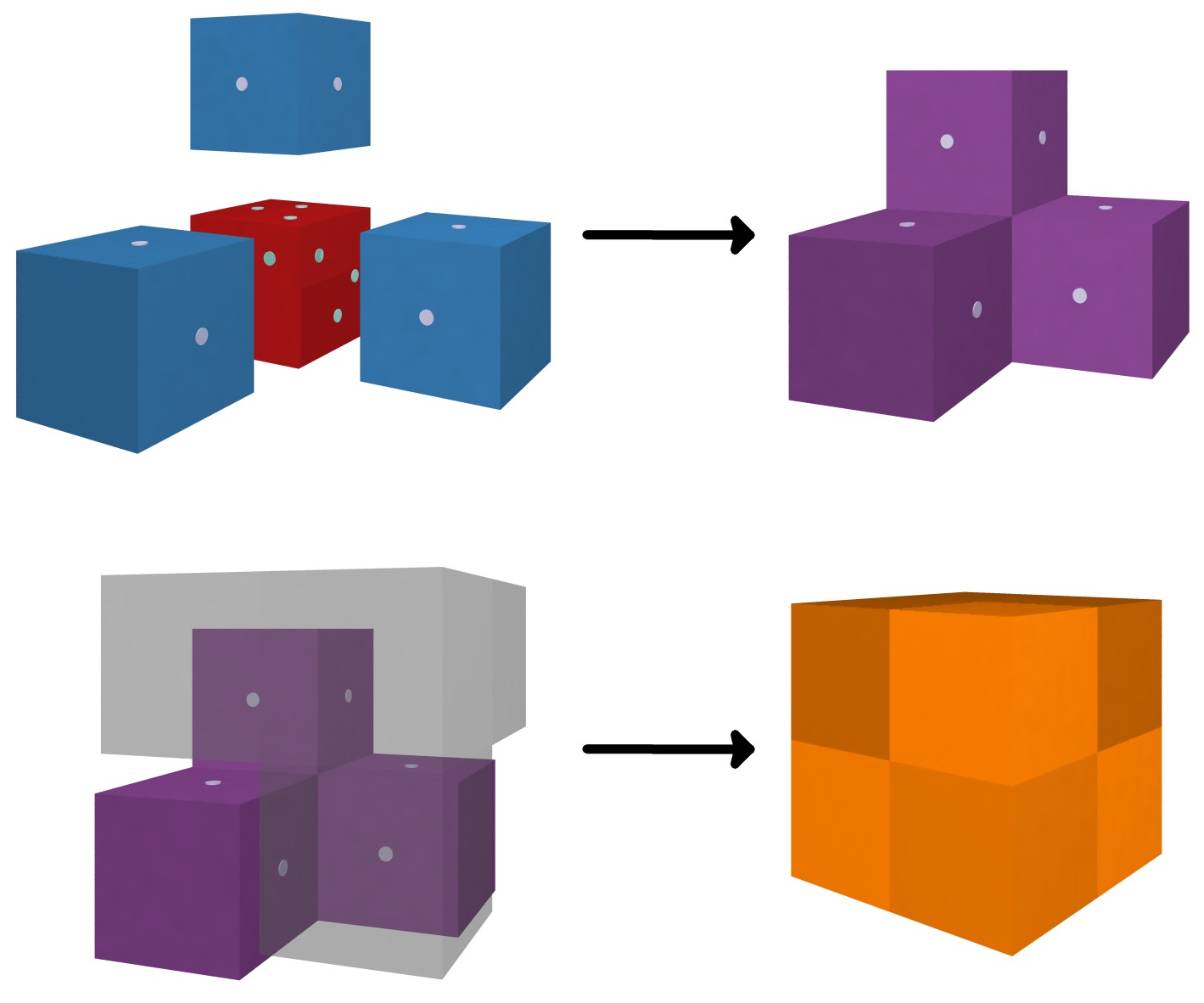}
  \caption{
    The intended hierarchical assembly pathway of model E.  Particle E1 (red)
    binds to the three E2 monomers (blue) to form a tetramer tripod (purple). Two
    tetramers may then link to form the octamer target (orange).
  }
  \label{fig:hier3_assembly}
\end{figure}

An alternative hierarchical model E employs a two-step process in which two
tripod-shaped fragments form and then interlock to make the target as
illustrated in \autoref{fig:hier3_assembly}.  Two particle types are again
required (\autoref{fig:models}, bottom row) but in this model they must be
combined in the ratio 1:3.  A three-letter alphabet of patches is required to
steer the model towards the anticipated pathway.  Patches on the E1
monomers, which form the apex of a tripod, bind exclusively to the patches
arranged in a triangle on one face of the E2 monomer.  The E1 patches do
not bind to each other, since this would make them identical to the
one-component model C, and neither do the corresponding patches on the E2
particles.  The third type of patch appears in the center of two adjacent
faces of the E2 monomer.  These patches on different E2 monomers bind exclusively
amongst themselves.  Six pairs of these patches come together when two
tripods bind correctly.

\begin{figure}[tbp]
  \centering
  \includegraphics[width=\singlefigure]{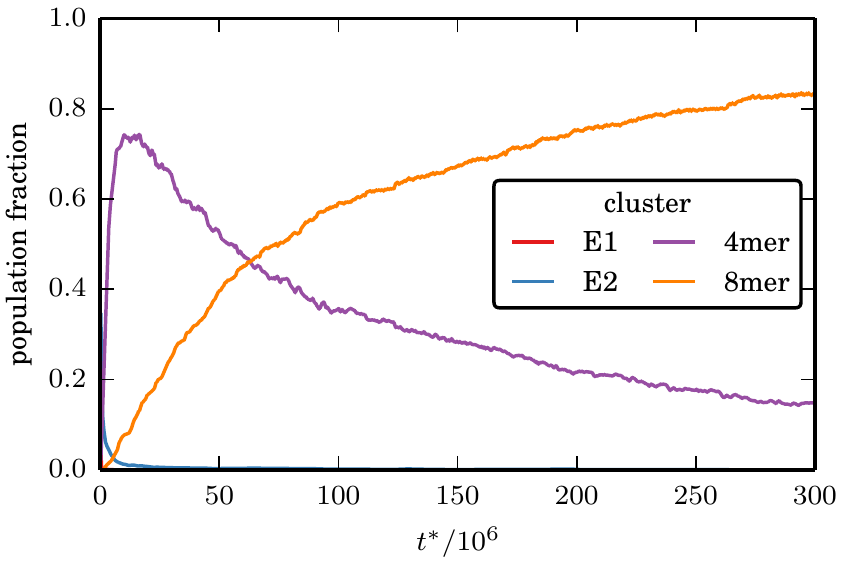}
  \caption{
    The population fraction of the target and correct fragments as a function of
    time for the hierarchical model E at $\Tstar = 0.044$.  \green{E1 monomers are
    rapidly incorporated into clusters and the corresponding line on the plot is
    barely visible.}
    Note the difference
    in the time scale compared with the other models, in particular the long
    time taken for tetramers to assemble into the octamer target.
  }
  \label{fig:hier3_assembly_44}
\end{figure}

\begin{figure}[tbp]
  \centering
  \includegraphics[width=\singlefigure]{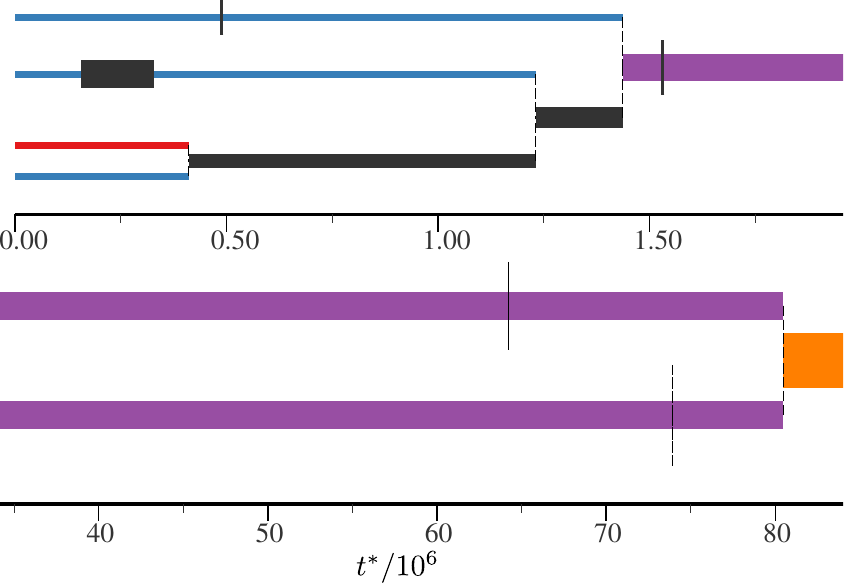}
  \caption{
    Example cluster histories for a single tetramer (top) and the formation of an
    octamer from tetramers on a much longer time scale (bottom) in hierarchical
    model E.
  }
  \label{fig:hier3_history}
\end{figure}

The two steps in the assembly of model E occur on well separated time scales.
\autoref{fig:hier3_assembly_44} shows that a large intermediate population
of tetramer tripods rapidly assembles from monomers, but complete targets only
emerge slowly.  The typical target history shown in \autoref{fig:hier3_history}
illustrates the sequential addition of monomers to make a tripod, which then
undergoes a long and uneventful trajectory before finally pairing up to complete
the target.
\par
Despite the substantial energetic reward for the second step of
assembly, there is a stringent steric requirement on the orientations with
which two tripods approach each other.  The purple trace in \autoref{fig:threshold}
shows that model E is mostly slower than the other hierarchical scheme D. However,
model E does have an advantage at low temperatures, where successful assembly persists
into the region in which models C and D have become very slow.  Success at low
temperature indicates the ability to avoid kinetic traps.  Model E may benefit from
taking place in only two steps, which allows a wider energetic separation between
the interactions that are promoted and suppressed in the first step.  The intermediate
is also relatively inert, and misaligned encounters are unlikely to lead to
significant interactions.  Hence, there is an absence of kinetic traps that does
become advantageous at low temperature.

\subsection{Addressable assembly}
\label{sec:dna}

The final assembly strategy that we simulate is the fully addressable limit
of DNA bricks, where all components of the target are distinct.  Model F
therefore consists of a mixture of eight different particles for the case of
our octamer target.
To make a point of contact with the other strategies, we use the same pattern
of patches as in the most efficient design so far, the sequential model C
(\autoref{fig:models}, top right).
However, each patterned face of each particle type in model F has a different patch
type, making a 24-letter alphabet.  Building block F1, for example, has three
faces that bind exclusively to particular faces on building blocks F2, F3 and F4,
respectively.  Continuing this scheme of pairwise interactions removes
all interactions that do not appear in the target structure.  All interacting
pairs of patches have the same strength, and the energy of the target is
therefore the same as in model C.
\par
\green{
The specificity in model F rules
out many of the fragments that would be incompatible with the final structure and
could act as traps.  However, the requirement of neighboring particles to be of a
particular species also removes a large number of fragments and targets that have the right
geometry but involve a combination of particle types that are not complementary in the
addressable design.
This labeling of particles greatly reduces the number of paths by which the target may
be built and amounts
to an entropic disadvantage that shifts optimal assembly of model F to lower
temperatures than model C, as shown in \autoref{fig:threshold}.
\par
The reduced number of paths generally also makes model F slower to assemble.} In the
one-component model C, an encounter between any two monomers can initiate assembly
and there are nine combinations of faces on the two particles that may bind correctly.
In contrast, a building block in the addressable case F must diffuse until it meets
one of the three other species with which it can bind, and binding may only occur
through one of the 36 possible combinations of faces.
\par
The addressable model is less robust with respect to temperature changes
than the other models examined so far.\cite{Reinhardt14a}
Despite the exclusion of incorrect
fragments, a thermodynamic yield of target at low temperatures can still be
prevented by the formation of multiple partially completed structures that starve the
system of the building blocks required to complete any one target.
At high temperature, assembly suffers from
the need for the right combination of building blocks to encounter each other
within a time frame shorter than the lifetime of a transient partially completed
structure.  \green{This need for a rare fluctuation is consistent with a recent
analysis showing that addressable assembly proceeds by a
non-classical nucleation process.\cite{Jacobs15b}}
\par
\begin{figure}[tbp]
  \centering
  \includegraphics[width=\singlefigure]{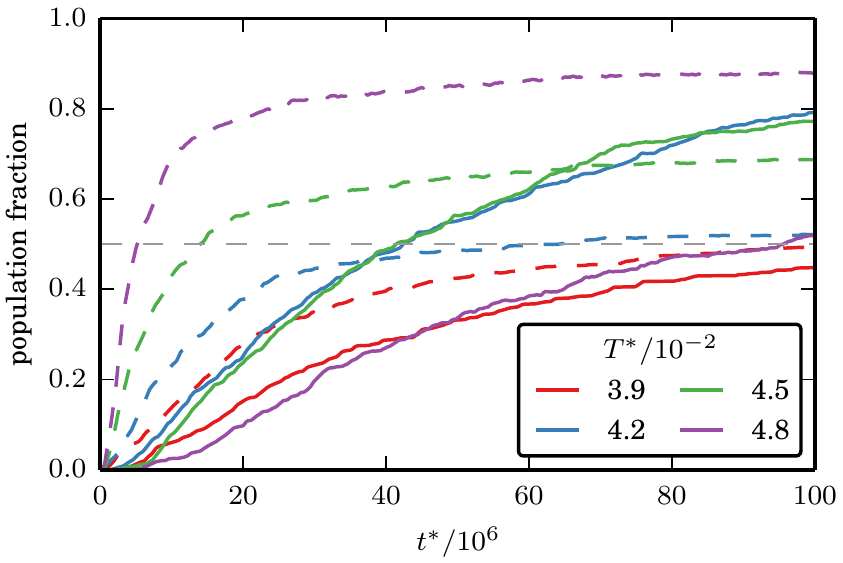}
  \caption{
    Comparison of the population fraction of the target structure as a function
    of time at four temperatures for both the addressable model F (solid lines)
    and the single-component model C (dashed lines).  The dashed horizontal line
    is the 50\% threshold at which $t_{1/2}^*$ is defined.
  }
  \label{fig:dna_torsion_comparison}
\end{figure}
\begin{figure}[tbp]
  \centering
  \includegraphics[width=\singlefigure]{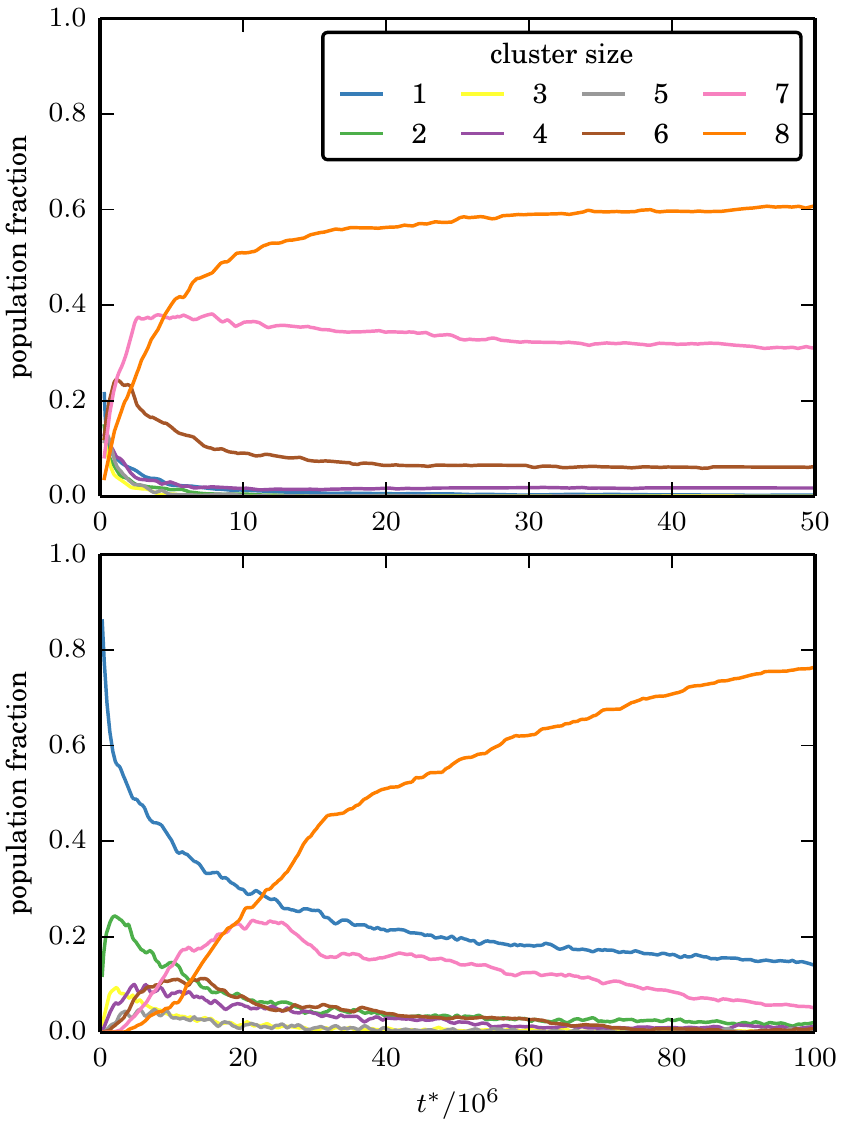}
  \caption{
\green{
Populations of fragments of 1 to 8 building blocks during assembly at reduced
temperature $T^*=0.045$ for the one-component model C (upper panel) and
addressable model F (lower panel).
}
  }
  \label{fig:fragsize}
\end{figure}

A more detailed comparison between the addressable model F and the
one-component model C shows that patience can reward the addressable strategy.
\autoref{fig:dna_torsion_comparison} shows that,
at some temperatures, the yield of target in model F rises slowly but steadily
beyond the point where model C slows down dramatically.  Hence, if the threshold
at which the assembly time is recorded for \autoref{fig:threshold} were raised
above 50\% yield (horizontal line in \autoref{fig:dna_torsion_comparison}),
the comparison would put the addressable strategy in a more favorable light.
\green{
\autoref{fig:fragsize} shows the contrasting evolution of fragment populations
in the two cases
at an intermediate temperature.  The upper panel shows that monomers in model
C are almost instantly incorporated into six- and seven-membered clusters that
then struggle to be completed due to the lack of monomers.  In the lower panel,
we see that the decay of monomers is much more gradual.  Although a low background
population of sizeable fragments arises, these partially completed structures can systematically
be completed, causing the yield of target structures to continue rising at times
when model C is virtually stuck.
}

\section{Concluding remarks}

For the small, highly symmetrical octamer target used in this work, the fastest and
most robust strategy for self-assembly was the sequential addition of identical
building blocks with a single type of interaction site.
The minimal viable design for the building block
(model B) has two patches on each of its interacting
faces to implement an effective torsional constraint that suppresses amorphous aggregation.
Additional patches can be used to improve the efficiency and reliability of self-assembly
somewhat further (model C) at the expense of greater complexity of building block.
The importance of a torsional potential has been noted in more coarse-grained
models.\cite{wilber_monodisperse_2009,Villar09a}
By implementing angular restrictions on binding
explicitly through a pattern of countable patches, \green{rather than by an
implicit effect built into the potential,} our model of patchy cubes provides
a way of quantifying the information required to implement a particular type of interaction.
\par
Although the octamer target does not have an intrinsically hierarchical structure, it is
nevertheless possible to envisage hierarchical pathways to its self-assembly.
Controlling not only
the final structure but also the assembly pathway is one of the goals of fully
programmable self-assembly.\cite{Rogers15a}  \green{Using two-component mixtures,}
we have devised a three-step (model D)
and a two-step (model E) scheme that promote hierarchical assembly by arranging for
each step to become energetically favorable only when the previous step has been
completed.\cite{Villar09a,Levy08a}
Evidence that these systems indeed assemble hierarchically
comes from history diagrams for individual
trajectories.  These diagrams trace the components of completed targets back through
time to see what types of aggregates the components belonged to as assembly progressed.
The analysis also revealed a change in mechanism from a budding process in amorphous
aggregates at low temperatures to a more orderly growth process at the optimal temperature.
\par
\green{Compared to the sequential addition of particles to make the octameric
target, hierarchical assembly requires a greater complexity of system, which
may take the form of differently patterned interfaces, a multi-letter alphabet of
interactions or a mixture of particle types.  However,}
for the octamer target, the hierarchical paths
held few advantages over the minimal model, despite their greater complexity.
It may be argued that a highly symmetrical target with no intrinsic modular structure
is unlikely to benefit from a hierarchical assembly strategy.  Nevertheless,
the benefits of multi-step assembly have also been called into question for systems
that do have a hierarchical arrangement of building blocks.\cite{Haxton13a}
\par
The most information-rich design tested in this work was the fully addressable limit
of eight particle species with exclusive interactions between faces that are adjacent
in the target structure.  The strategy of eliminating interactions between sites
that are not in contact in the final structure is akin to G{\=o} models for
proteins,\cite{Ueda78a} in which non-native interactions between amino acids are set
to zero to improve the folding properties.\cite{Miller99b}  Unlike proteins,
however, the components of a DNA brick
system\cite{Ke12a} are not connected by a backbone and must encounter each other by
diffusion.  The need to find specific binding partners generally leads to slower assembly
and can even increase the problem of monomer starvation at low temperatures,
since a specific combination of particles is needed to complete a target, and
the formation of partial fragments can prevent completion of any given partial
structure.  We note that self-assembly simulations that include only enough building
blocks for one complete copy of the target do not capture this important source of
frustration in addressable systems.  Although the addressable version of our model
was slower to assemble and less tolerant of changes in temperature, it
\green{sometimes produced a higher yield at sufficiently long times because,
at low temperatures there was less competition from erroneous structures, while at
moderate temperatures the slower assembly process helped to avoid premature consumption
of building blocks.}
\par
We have taken a largely intuitive approach to the building block designs explored in
this work, attempting to place attractive patches at locations that can reasonably
be expected to promote the
target structure while avoiding obvious unintended structures by careful choice of
patch spacing.  Our designs are almost certainly not optimal even within the constraints
imposed by the different classes of design strategy.  One approach to improving the
designs would be a genetic or evolutionary algorithm that is driven by a fitness function
based on speed of assembly or yield of target.  It would be interesting to see
whether such algorithms would automatically favor one or other of the broad classes of
design strategy, or even arrive at strategies that have not been envisaged here.
There is clearly scope for investigating more complex targets than the symmetrical
octamer used in this paper, and it is highly likely that the different self-assembly
strategies will reach their limits for different sizes and types of target structures.
The patchy cube model, in conjunction with the hybrid Monte Carlo
algorithm presented here, should be a versatile workhorse for
testing different strategies for self-assembly given the constraints
applicable to a particular type of experimental building block.

\section*{Acknowledgements}

The authors thank Dr Laura Filion for benchmark results to test our code for
unpatterned hard cubes, and Dr Thomas Ouldridge for insight into the VMMC
algorithm.  M.A.M.~is grateful for support from EPSRC Programme Grant EP/I001352/1.

\begin{NoHyper}
\end{NoHyper}
\end{document}